\documentclass{aa}
\usepackage[varg]{txfonts}
\usepackage{natbib,twoopt}
\usepackage[breaklinks=true]{hyperref} %% to avoid \citeads line fills
\bibpunct{(}{)}{;}{a}{}{,} %% natbib format for A&A and ApJ
\usepackage{graphicx}
\usepackage{subcaption}	% for subfigures
\usepackage{amsmath}	% for case environment
\usepackage{float}
\usepackage{color}
\usepackage{verbatim}
\usepackage[normalem]{ulem}
\usepackage{rotating}		% sidewaystable
\usepackage{array}

\begin{document}
	
	\title{The stellar corona-chromosphere connection}
\subtitle{
A comprehensive study of X-ray and Ca II IRT fluxes from eROSITA and Gaia
} 
	\author{S. Freund\inst{1,2} \and S. Czesla\inst{3,2} \and B. Fuhrmeister\inst{3,2} \and P. Predehl\inst{1} \and J. Robrade\inst{2} \and P.C. Schneider\inst{2} \and J.H.M.M. Schmitt\inst{2}}
	\institute{Max-Planck-Institut für extraterrestrische Physik, Gießenbachstraße 1, 85748 Garching Germany
		\email{sfreund@mpe.mpg.de}\label{inst1}%
		\and
		Hamburger Sternwarte, Universit\"at Hamburg, 21029 Hamburg, Germany\label{inst2}%
		\and
		Th\"uringer Landessternwarte Tautenburg, Sternwarte 5, D-07778 Tautenburg, Germany\label{inst3} %
  }

  \date{Received 08 July 2024 / Accepted 29 March 2025}
  
	\abstract{Stellar activity can be observed at different wavelengths in a variety of different activity indicators. We investigated the correlation between coronal and chromospheric emissions by combining X-ray data from stars detected in the \textit{eROSITA} all-sky surveys (eRASS1 and eRASS:5) with Ca~II infrared triplet (IRT) activity indices as published in the third \textit{Gaia} data release (\textit{Gaia} DR3). We specifically studied 24\,300 and 43\,200 stellar sources with reliable Ca II IRT measurement and X-ray detection in eRASS1 and eRASS:5, which is by far the largest stellar sample available so far. The largest detection fraction is obtained for highly active sources and stars of a late spectral type, while F-type and less active stars (as measured in the Ca~II IRT) remain mostly undetected in X-rays. Also, the correlation is the strongest for late-type sources, while F-type stars show a rather weak correlation between the X-ray to bolometric flux ratio and the Ca~II IRT activity index. The relation between the X-ray and Ca~II IRT surface fluxes changes with the fractional X-ray flux without showing two separated branches as described in previous studies. For fast rotators, both activity indicators saturate at a similar Rossby number and the X-ray to bolometric flux ratio decreases faster than the IRT index for slower rotating stars. As a consequence, the ratio between X-ray and IRT fluxes is constant in the saturation regime and decreases for slow rotators.}
 
	\maketitle

	\section{Introduction}
	\label{sec: introduction}
    Magnetic activity of late-type stars reveals itself in different phenomena that originate from different layers of the stellar atmosphere and can be observed at different wavelengths. In the optical, the chromospheric Ca II H\&K lines have been studied extensively \citep{wilson78,linsky79,baliunas95,hall09,hempelmann16} and also the H$\alpha$ line \citep{he19} and the Ca II infrared triplet (IRT) lines are well known activity indicators. X-rays emitted from the magnetically heated coronae provide another opportunity to study stellar activity. \citet{martin17} compared Ca II IRT with Ca II H\&K for more than 2\,000 spectra of almost 100 stars obtained by the \textit{TIGRE} telescope. Also, the correlation between the activity indicators derived from the Ca II H\&K lines and X-rays have been intensively investigated \citep{schrijver83,mittag18,fuhrmeister22} but comparisons between the Ca II IRT and X-rays are far less numerous. \citet{martinez11} analyzed flux-flux relationships including basal corrected Ca II IRT emission for a sample of 298 F- to M-type dwarfs with 243 of them being detected in X-rays in the \textit{ROSAT} all-sky survey (RASS). In addition, \citet{stelzer13} present a sample of 24 non-accreting pre-main sequence stars for which Ca II IRT and other chromospheric fluxes are estimated from X-shooter/VLT spectra; X-ray fluxes are available for 19 of these sources.

    Ca~II IRT activity measurements for a large number of stars have now become available from the \textit{Gaia} mission \citep{GaiaMission} launched in 2013. In addition to the highly accurate positions, parallaxes, and photometry, high resolution spectra are obtained with the Radial Velocity Spectrometer (RVS) for the brightest \textit{Gaia} sources. With an effective wavelength range of 846 - 870~nm these spectra do cover the Ca~II IRT with a resolution of $R\approx 11\,500$, and \citet{lanzafame23} used these spectra to estimate Ca~II IRT based activity indices for about 2 million \textit{Gaia} DR3 sources. 

    On the X-ray side, the deepest all-sky survey at soft X-ray wavelengths is performed by the \textit{eROSITA} X-ray telescope \citep{predehl21} onboard the Spectrum-Roentgen-Gamma mission \citep[SRG;][]{sunyaev21}. The X-ray sources detected during the first \textit{eROSITA} all-sky survey (eRASS1) carried out between December 2019 and June 2020 are presented by \citet{merloni24}. The main eRASS1 source catalog contains about 900\,000 point-like sources detected in the 0.2~--~2.3~keV band and located in the western Galactic hemisphere, with nearly 140\,000 of them expected to be produced by stellar coronal emission \citep{freund24}. After eRASS1, the survey was continued until February 2022 when four all-sky surveys were finished and the fifth survey was partially completed revealing an even larger sample of stellar X-ray sources.

    As evidenced by the Sun, chromospheric and coronal emissions are not homogeneously produced over the stellar surface, instead the bulk of the emission is concentrated in active regions. The fraction of the stellar surface covered by active regions and producing most of the activity related emission is often described in terms of a filling factor, which might vary between the corona and the chromosphere. Measurements at X-ray and optical wavelengths cannot spatially resolve stellar surfaces,  rather the X-ray and IRT luminosities measured by \textit{eROSITA} and \textit{Gaia} are integrated over the whole surface and any reported surface fluxes are then surface averaged mean fluxes. 

    The combination of \textit{Gaia} and \textit{eROSITA} data provide a sample of stars with available Ca~II IRT and X-ray measurements of unprecedented size and allows us to study in detail the correlation between the two activity indicators. This paper is structured as follows: we describe the data being used in Sect.~\ref{sec: data and measuremnts}. In Sect.~\ref{sec: X-ray detections} we specify the fraction of sources with Ca~II IRT activity index also detected in X-rays. We discuss various aspects of the correlation between the X-ray and Ca II IRT emission in Sect.~\ref{sec: Correlation between chromospheric and X-ray activity}, and finally draw our conclusions in Sect.~\ref{sec: Conclusion}.

	\section{Data and measurements}
    \label{sec: data and measuremnts}
	\subsection{\textit{Gaia} DR3 and Ca II IRT measurements}
	The data taken during the first 34 months of the \textit{Gaia} mission \citep{GaiaMission} are published in the third \textit{Gaia} data release \citep[\textit{Gaia} DR3;][]{GaiaEDR3, GaiaDR3}. It contains sub-milliarcsecond accuracy positions for 1.8 billion sources brighter than 21$^{st}$ magnitude. Furthermore, photometry in three bands (G, BP, and RP) as well as parallaxes and proper motions are provided for 1.5 billion sources. \textit{Gaia} DR3 is expected to be highly complete in the range $7<G<19$~mag with some of the brightest and faintest sources and objects in crowded regions missing from the catalog. 
	
    In addition to the photometry, high resolution spectra from the RVS and low resolution spectra from the BP and RP prism photometers covering the wavelength ranges of 330 - 680~nm and 640 - 1050~nm with a resolution between 30 and 100 in $\lambda/\Delta\lambda$ are also obtained as part of \textit{Gaia} DR3. The General Stellar Parametrizer from Photometry (GSP-Phot) estimated stellar parameters, e.g., effective temperature, surface gravity, metallicity, radius, reddening, and extinction, from the BP and RP spectra for about 500 million sources brighter than $G=19$~mag. The effective temperature and metallicity were also estimated from the RVS spectroscopy (GSP-Spec) for about 6 million sources brighter than $G=14$~mag. 
	
	To derive Ca~II IRT activity indices, \citet{lanzafame23} used theoretical photospheric spectra for a linear interpolation over a grid of MARCS synthetic spectra adopting the stellar parameters from GSP-Spec whenever available, and otherwise from GSP-Phot. For the synthetic spectra, they assumed local thermodynamic and radiative equilibrium and applied rotational broadening. The activity index $\alpha$ was derived from the ratio between the observed and the template spectrum around the cores of each of the triplet lines with the mean of the three values taken as activity index. The so-obtained activity index is close to zero (or even negative) for inactive stars and positive for active stars, however, it is not suitable for a comparison of the activity of stars with different stellar parameters, i.e., effective temperature, surface gravity, and metallicity. Therefore, \citet{lanzafame23} defined a $R^\prime_\mathrm{IRT}$-index and provide a conversion between the activity index $\alpha$ and the $R^\prime_\mathrm{IRT}$-index as a function of the effective temperature and metallicity. The $R^\prime_\mathrm{IRT}$-index is connected with the surface chromospheric flux $\mathcal{F}_\mathrm{IRT}$ and the observed flux $F_\mathrm{IRT}$ of the IRT lines through the bolometric flux $F_\mathrm{bol}$ \citep[compare Equation~6 of][]{lanzafame23}
	\begin{equation}
		R^\prime_\mathrm{IRT} = \frac{\mathcal{F}_\mathrm{IRT}}{\sigma T^4_\mathrm{eff}}
		  = \frac{\mathcal{F}_\mathrm{IRT} A}{\sigma T^4_\mathrm{eff} A}
		  = \frac{L_\mathrm{IRT}}{L_\mathrm{bol}}
		  = \frac{F_\mathrm{IRT}}{F_\mathrm{bol}},
		\label{equ: flux IRT}
	\end{equation}
	where $T_\mathrm{eff}$ is the effective temperature, $A$  the surface of the star, and $L_\mathrm{bol}$ and $L_\mathrm{IRT}$ are the bolometric luminosity and the luminosity of the IRT lines, respectively. We emphasize that $\mathcal{F}_\mathrm{IRT}$, $F_\mathrm{IRT}$, $F_\mathrm{bol}$, $L_\mathrm{IRT}$, and $L_\mathrm{bol}$ are the observed mean fluxes and total luminosities.
	
	\subsection{\textit{eROSITA} observations and stellar identification \label{sect:ero}}
 
	Similar to the method described in \citet{merloni24} for eRASS1, all data collected during the \textit{eROSITA} all-sky surveys were combined and  the source detection was applied to this merged data set. The resulting source catalog (eRASS:5) in the western Galactic hemisphere of the sky is available to the German \textit{eROSITA} consortium, and we refer in the following to the catalog version 230619 obtained by eSASS version c020 \citep{brunner22}. eRASS:5 has a mean vignetting corrected exposure time of about 950~s. However, the survey exposure time is not constant, varying between the ecliptic equator and poles between about 300~s and several thousand seconds due to the \textit{eROSITA} scanning law. Applying an energy conversion factor (ECF) of $8.5\times 10^{-13}$~erg~cm$^{-2}$~cnt$^{-1}$ as appropriate for most coronal sources, a mean sensitivity of about $2.0\times 10^{-14}$~erg~cm$^{-2}$~s$^{-1}$ results in the 0.2~--~2.3~keV band for eRASS:5 (compared to $5.5\times 10^{-14}$~erg~cm$^{-2}$~s$^{-1}$ for eRASS1) with substantially higher sensitivities being reached in the vicinity of the southern ecliptic pole. The positional uncertainty of the eRASS:5 sources is on average about 4.3~arcsec, with the positional accuracy generally increasing for brighter X-ray sources due to better photon statistics.

    The identification of the coronal content for the eRASS1 catalog is described by \citet{freund24} and we applied the same method for eRASS:5.
	Specifically, we performed a crossmatch of the eRASS:5 sources with eligible counterparts from \textit{Gaia} DR3, i.e., sources with flux measurements in all three \textit{Gaia} bands, an accurate parallax ($\frac{\pi}{\sigma_\pi}>3$), and brighter than $G=19$~mag. Geometric properties as the angular separation between the X-ray and \textit{Gaia} position and the eRASS:5 positional uncertainty were considered in combination with additional properties, namely the \texttt{BP-RP} color, X-ray over bolometric flux, and the distances of the counterparts, within a Bayesian framework to identify the correct association.
	As a result, we obtain 408\,000 likely coronal sources out of 2.9 million point-like X-ray sources and expect these identifications to be complete and reliable to about 89\% based on the coronal probabilities estimated for every eRASS:5 source (compared to 138\,800 coronal eRASS1 identifications with a completeness and reliability of 91.5\%).

	\section{Source selection and statistics of X-ray detections}
    \label{sec: X-ray detections}

    We considered \textit{Gaia} Ca~II IRT activity indices only for eligible coronal counterparts according to our definition (Sect.~\ref{sect:ero}) and sources located in western Galactic sky hemisphere. Furthermore, we excluded sources with negative activity index values and those objects for which the derived activity index does not exceed three times its estimated uncertainty as we consider the activity of these sources to have remained undetected with the Gaia RVS spectra. 
    This leaves us with 556\,700 sources with a measured Ca~II IRT activity index and 34\,300 and 70\,600 of them are 
    identified as likely coronal counterparts of an eRAS1 and eRASS:5 source, respectively.
	A further set of 13\,100 sources with a Ca~II IRT measurement are located within the formal $5\sigma$ of the eRASS:5 positional uncertainty, yet these sources have rather low matching probabilities and are -- in their majority -- more likely chance alignments.
	However, our identification procedure does not consider the Ca~II IRT activity and the matching probability of some sources with high Ca II IRT activity might increase if this additional information was considered. To obtain the most reliable association, we nonetheless restricted the analysis in this paper to the likely identifications obtained with the method described in \citet{freund24}. 
	
	For the sources with both Ca II IRT activity index and eRASS:5 identification, we applied Equation~7 and Table~1 of \citet{lanzafame23} to estimate the value of $\log(R^\prime_\mathrm{IRT})$ by adopting the same stellar parameters used by \citet{lanzafame23}. 
    For a few sources, this equation cannot be used because the effective temperature and/or the metallicity are not available in \textit{Gaia} DR3.
    Thus, we are left with 70\,300 sources with a valid $\log(R^\prime_\mathrm{IRT})$-index and a good eRASS:5 identification.  
    
	\begin{figure}[t]
		\includegraphics[width=\hsize]{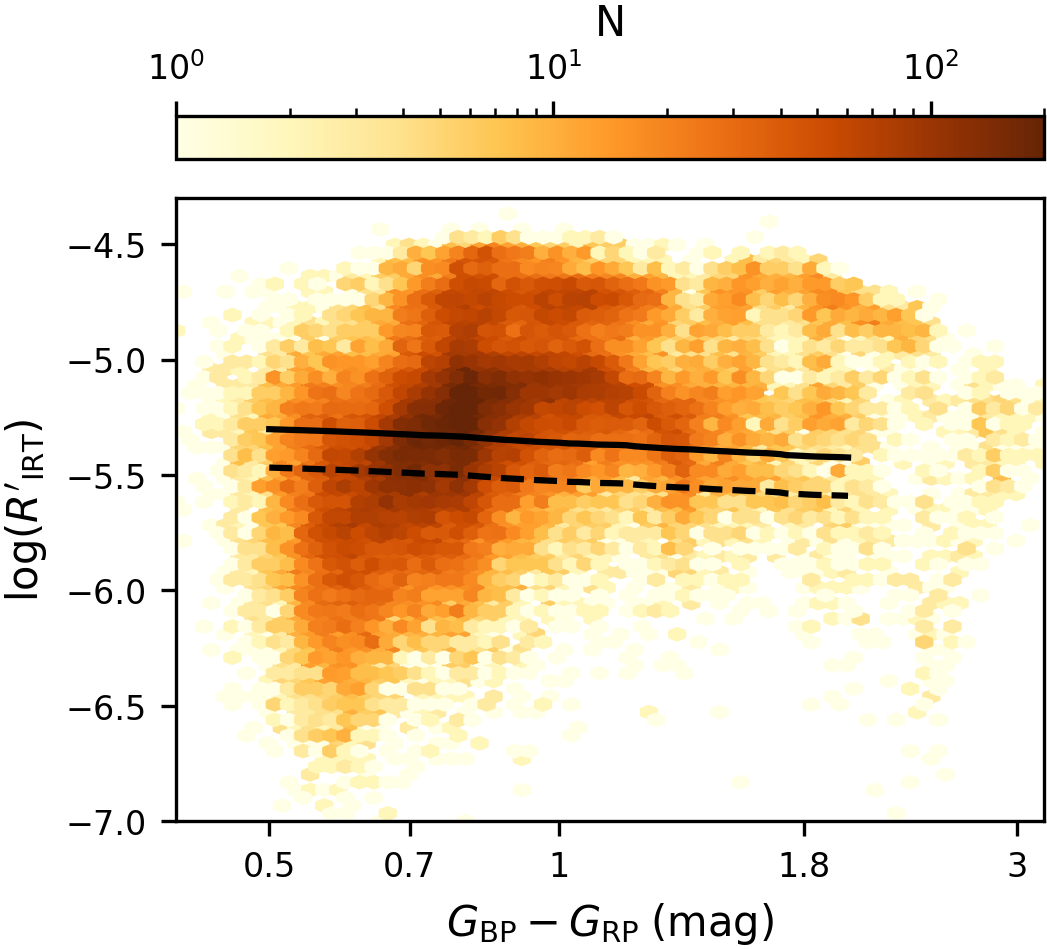}
		\caption{Distribution of the $\log(R^\prime_\mathrm{IRT})$-index as a function of the $G_\mathrm{BP}-G_\mathrm{RP}$ color for the \textit{Gaia} DR3 sources with a reliable eRASS:5 identification. The color scales with the number of sources within the bin. The black solid and dashed lines show the $\log(R^\prime_\mathrm{IRT})$-indices corresponding to the standard deviation of low active stars for GSP-Phot and GSP-Spec input as described in the \textit{Gaia} DR3 online documentation (see text for details).}
		\label{fig: basal IRT index}
	\end{figure} 

    In Fig.~\ref{fig: basal IRT index} we show the distribution of the $R^\prime_\mathrm{IRT}$-index as a function of the $G_\mathrm{BP}-G_\mathrm{RP}$ color\footnote{We use extinction corrected \textit{Gaia} magnitudes and reddening corrected colors throughout this paper adopting the values from \citet{Gaia_gspphot}} for our eROSITA X-ray detected sources. Although we excluded sources with activity indices of 
    low significance, the $R^\prime_\mathrm{IRT}$-index still reaches quite low values because the provided uncertainty of the activity index does not consider all sources of error. The quoted errors  typically have values of about $0.001$~nm, yet according to the \textit{Gaia}~DR3 online 
    documentation \citep{GaiaAP}, a sample of activity indices of low activity stars has a standard deviation of 0.015 and 0.022~nm for GSP-Spec and GSP-Phot input, which is an upper limit of the random errors. In Fig.~\ref{fig: basal IRT index} we show the $R^\prime_\mathrm{IRT}$-index corresponding 
    to these limits as dashed and solid black lines. Ca II IRT activity below this limit cannot be reliably detected 
    in \textit{Gaia}~DR3, hence we consider stars with $\log(R^\prime_\mathrm{IRT})<-5.4$ as inactive in the 
    following. This leaves us with 24\,300 eRASS1 and 43\,200 eRASS:5 X-ray detections out of 180\,700 sources with reliable activity indices.

	\begin{figure}[t]
		\includegraphics[width=\hsize]{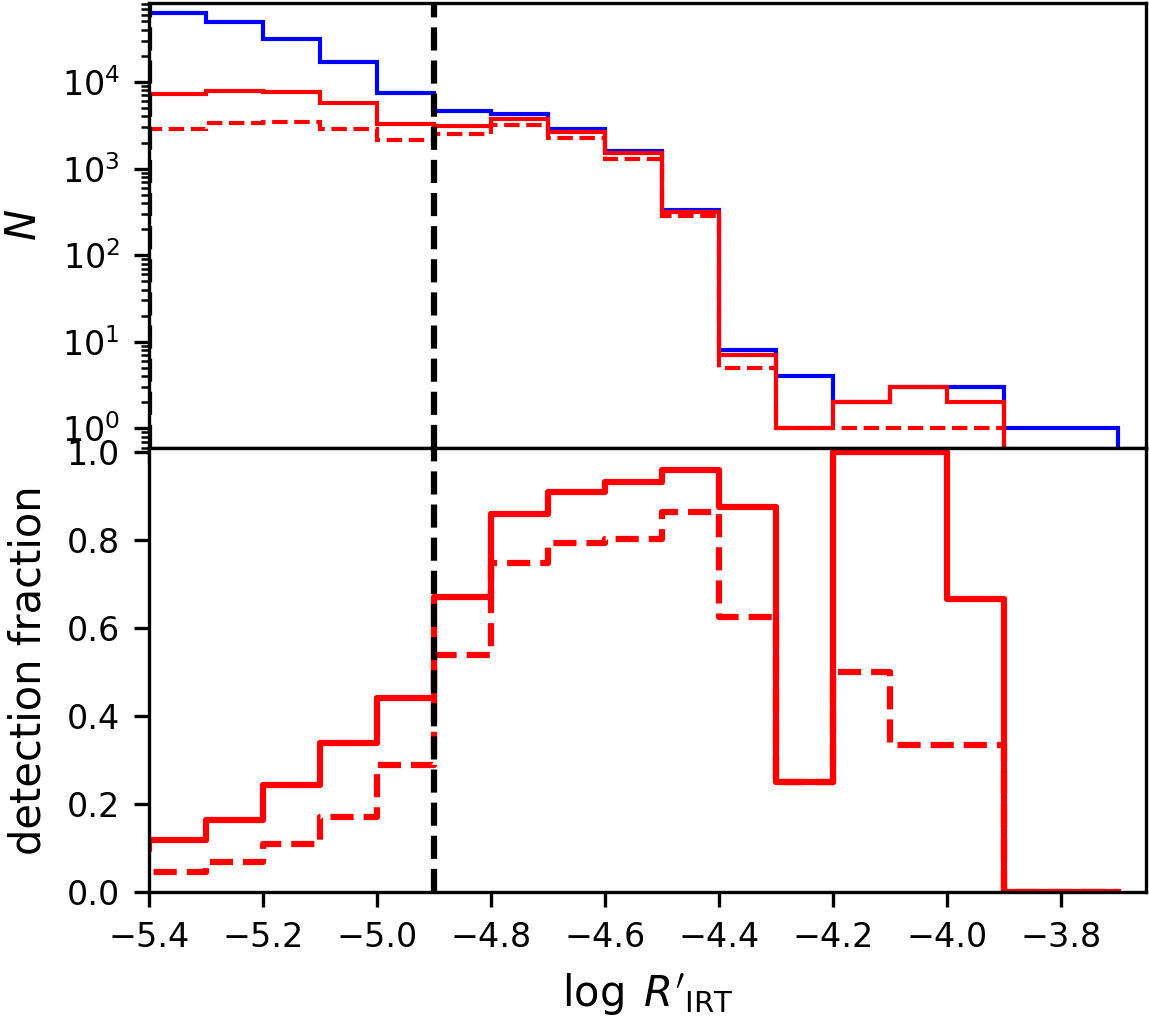}
		\caption{Distribution of $\log(R^\prime_\mathrm{IRT})$ for \textit{Gaia} DR3 sources. Upper panel: Total number of all sources (blue histogram) and those sources identified as eRASS1 and eRASS:5 X-ray emitters (red dashed and solid histogram).  Lower Panel:  X-ray detection fraction vs. $R^\prime_\mathrm{IRT}$-index; the dashed line indicates the boundary between the HA and VHA groups (see text for details).}
		\label{fig: activity index log_R}
	\end{figure}  
	
	In Fig.~\ref{fig: activity index log_R} we compare the $R^\prime_\mathrm{IRT}$-index distribution for all \textit{Gaia} DR3 sources 
    and those sources also detected as X-ray sources in eRASS1 and eRASS:5. Fig.~\ref{fig: activity index log_R} shows that most of the sources with $\log(R^\prime_\mathrm{IRT}) > -5$ are detected in eRASS:5 and the detection fraction increases up to more than 90~\%.  However, the fraction of X-ray detected sources rapidly decreases for smaller $R^\prime_\mathrm{IRT}$-indices. Especially sources with a low $R^\prime_\mathrm{IRT}$-index remain undetected in eRASS1.
    Following \citet{lanzafame23} we divided our sample in different groups: the very high activity group (VHA; $\log(R^\prime_\mathrm{IRT})>-4.9$) and the high activity group (HA; $-5.4<\log(R^\prime_\mathrm{IRT})<-4.9$). 
    Lower activity values are usually not reliably detected in \textit{Gaia}~DR3 and these sources can be considered as inactive (IA; $\log(R^\prime_\mathrm{IRT})<-5.4$) so that no correlation with other activity indicators such as X-rays is expected within this group.
 
    In Table~\ref{tab: eRASS:5 detection fraction} we provide the total number of sources and the number of likely eRASS1 and eRASS:5 identifications in each of these groups separated by spectral type and luminosity class. Specifically, we divided our sample in giants (located more than 2~mag above the main sequence and with $G_\mathrm{BP}-G_\mathrm{RP}<1.84$~mag) and dwarfs of spectral type F ($G_\mathrm{BP}-G_\mathrm{RP}<0.78$~mag), G ($0.78\leq G_\mathrm{BP}-G_\mathrm{RP} < 0.98$~mag), K ($0.98\leq G_\mathrm{BP}-G_\mathrm{RP} < 1.84$~mag), and M ($G_\mathrm{BP}-G_\mathrm{RP}\geq 1.84$~mag). For the VHA sources, more than 80\% are detected as X-ray sources in eRASS:5 but this fraction decreases to 7.2\% for the inactive sources; only for stars classified as F-type we find a detection rate of only 54\% for the VHA sources.
    
	\begin{table*}[t]
		\caption{X-ray detection fraction}
		\label{tab: eRASS:5 detection fraction}
        \small
		\centering
        
		\begin{tabular}{l |*{3}{r} |*{3}{r} |*{3}{r}}
			\hline \hline
		       & \multicolumn{3}{c|}{VHA} & \multicolumn{3}{c|}{HA} & \multicolumn{3}{c}{IA} \\
             & \multicolumn{2}{c}{detected} & total & \multicolumn{2}{c}{detected} & total & \multicolumn{2}{c}{detected} & total \\
             & eRASS1 & eRASS:5 & & eRASS1 & eRASS:5 & & eRASS1 & eRASS:5 & \\
			\hline
			F-type & 1\,100 (39\%) & 1\,600 (54\%) & 2\,900 & 5\,000 (10\%) & 10\,300 (21\%) & 48\,200 & 5\,800 (4.3\%) & 15\,400 (11\%) & 134\,700 \\
            G-type & 3\,100 (77\%) & 3\,600 (90\%) & 4\,000 & 5\,000 (9.5\%) & 10\,900 (21\%) & 52\,700 & 1\,400 (1.3\%) & 4\,400 (4.0\%) & 109\,600 \\
            K-type & 3\,200 (85\%) & 3\,500 (92\%) & 3\,800 & 3\,000 (8.2\%) & 7\,400 (20\%) & 37\,100 & 710 (1.7\%) & 2\,300 (5.4\%) & 43\,100 \\
            M-type & 500 (88\%) & 520 (91\%) & 570 & 430 (21\%) & 620 (30\%) & 2\,100 & 220 (7.4\%) & 510 (17\%) & 3\,000 \\
            Giants & 1\,700 (66\%) & 2\,100 (82\%) & 2\,500 & 1\,300 (4.9\%) & 2\,700 (10\%) & 26\,900 & 1\,700 (2.1\%) & 4\,400 (5.2\%) & 84\,700 \\
            Total & 9\,600 (70\%) & 11\,300 (82\%) & 13\,800 & 14\,700 (8.8\%) & 31\,900 (19\%) & 167\,000 & 9\,800 (2.6\%) & 27\,000 (7.2\%) & 375\,100 \\ 
			\hline
		\end{tabular}
	\end{table*}
 
	\begin{figure*}[t]
		\centering
		\includegraphics[width=0.48\textwidth]{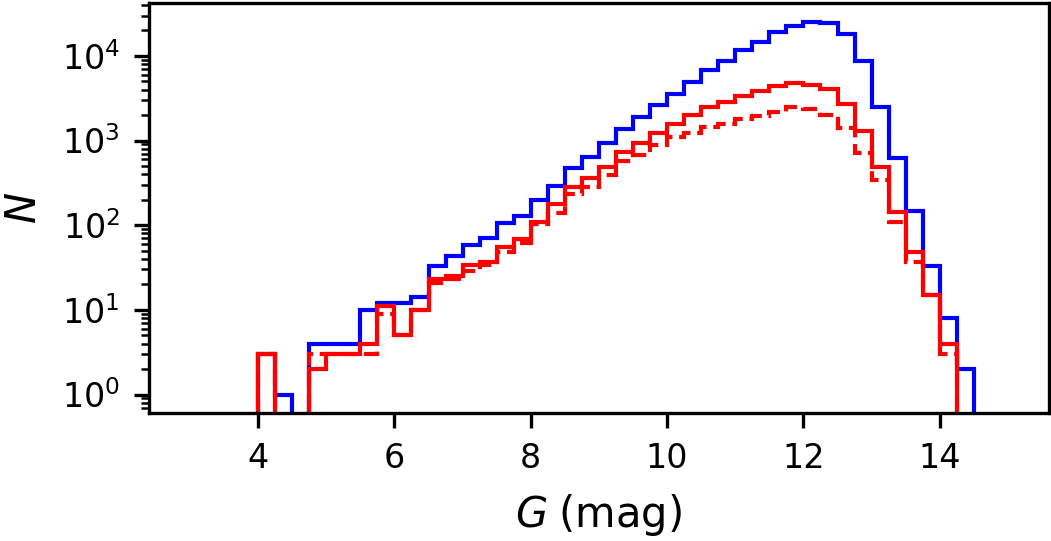}
        \hfill
		\includegraphics[width=0.48\textwidth]{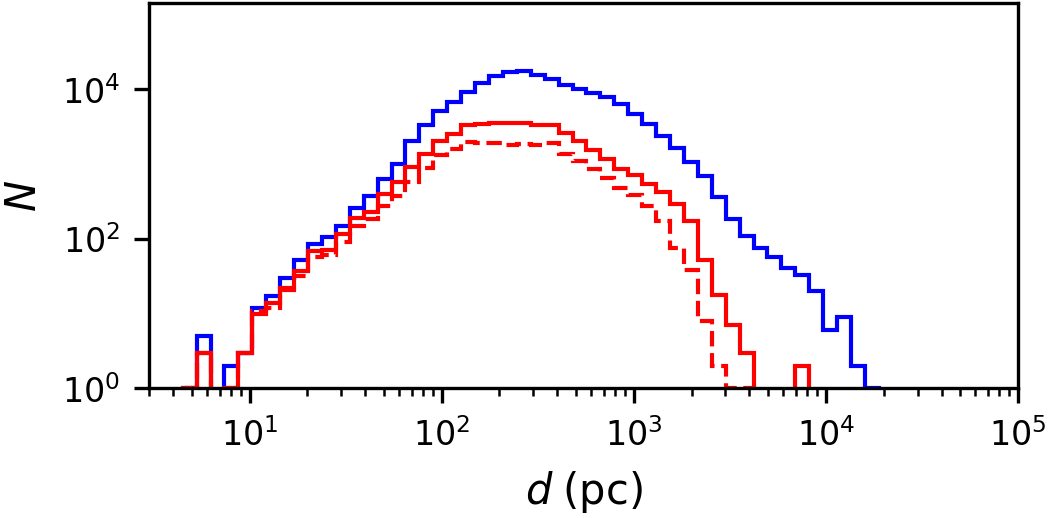}
        \par\bigskip
		\includegraphics[width=0.48\textwidth]{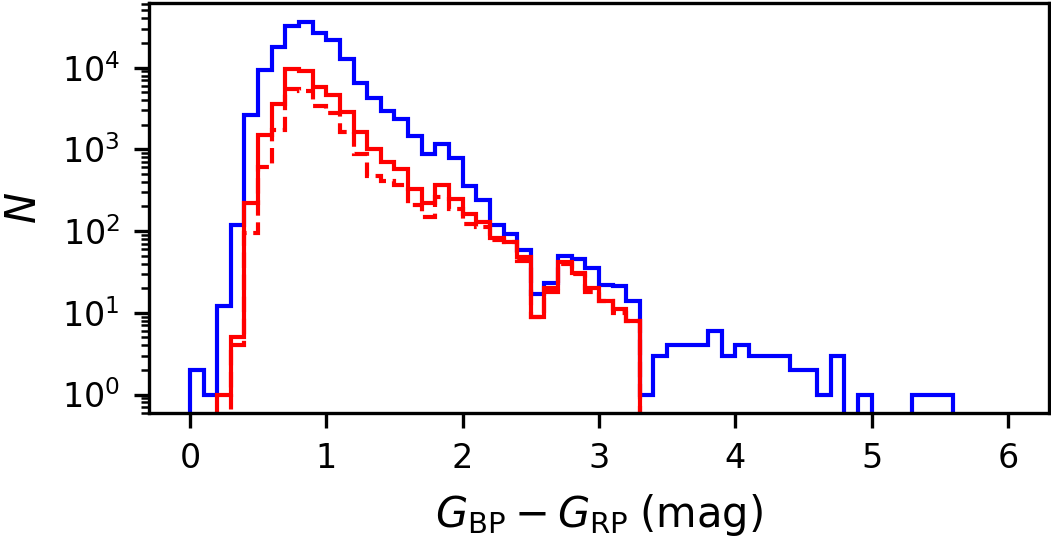}
        \hfill
		\includegraphics[width=0.48\textwidth]{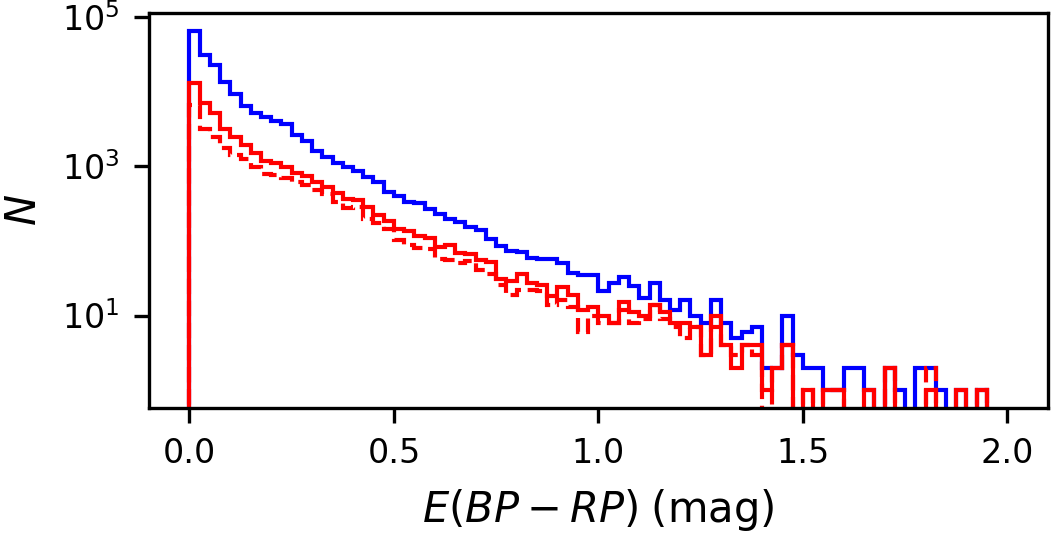}
		\caption{G-band magnitude, distance, $G_\mathrm{BP}-G_\mathrm{RP}$ color, and reddening distribution of the \textit{Gaia} DR3 sources with reliable activity index. The blue histogram shows the distribution for all and those sources identified as eRASS1 and eRASS:5 X-ray emitter are indicated by the red dashed and solid lines, respectively.}
		\label{fig: activity index properties comparison}
	\end{figure*}

	In Fig.~\ref{fig: activity index properties comparison}, we compare the G-band magnitudes, distances, $G_\mathrm{BP}-G_\mathrm{RP}$ colors and the reddening for all sources with reliable Ca~II IRT measurements with those also identified as eRASS1 and eRASS:5 X-ray emitters. In general, the sources detected in X-rays are somewhat brighter and closer to the Sun.  
	Furthermore, we detect more of the redder sources in X-rays but the reddest objects with $G_\mathrm{BP}-G_\mathrm{RP}$-values greater than about 3~mag remain undetected; an inspection of these sources revealed that they are typically giants with very little or no X-ray emission (see \cite{schmitt2024} for a detailed study). Since the majority of the sources in our sample is quite close to the Sun, the reddening is small for most objects, but extends to larger values for a few sources. The sample of eRASS1 detected sources has overall very similar properties although more of the sources at higher distances remain undetected.

    Since \citet{lanzafame23} estimated the Ca~II IRT activity index by subtracting photospheric model spectra, their 
    activity indices do include the basal flux. A basal flux contribution is well known for the Ca~II~H\&K lines \citep[e.g.][]{mittag13}. Together with conversions between the Ca II IRT and Ca II H\&K activity, e.g. from \citet{martin17} or \citet{lanzafame23}, the $R^\prime_\mathrm{IRT}$-index corresponding to the Ca II H\&K basal flux can be estimated. (We note that there are also direct basal flux estimates for the Ca~II~IRT lines \citep{martin17PhD}.) The resulting value of $\log(R^\prime_\mathrm{IRT})$ is expected between about $-6.0$ and $-5.5$ depending on the color and the adopted IRT-H\&K conversion, making its detection challenging with the current \textit{Gaia} data. However, more sensitive data, possibly contained in the next \textit{Gaia} data release, are likely to improve the detectability of the expected basal IRT flux contribution.

	\section{Correlation between chromospheric and X-ray activity}
    \label{sec: Correlation between chromospheric and X-ray activity}

\begin{figure}[t]
		\includegraphics[width=\hsize]{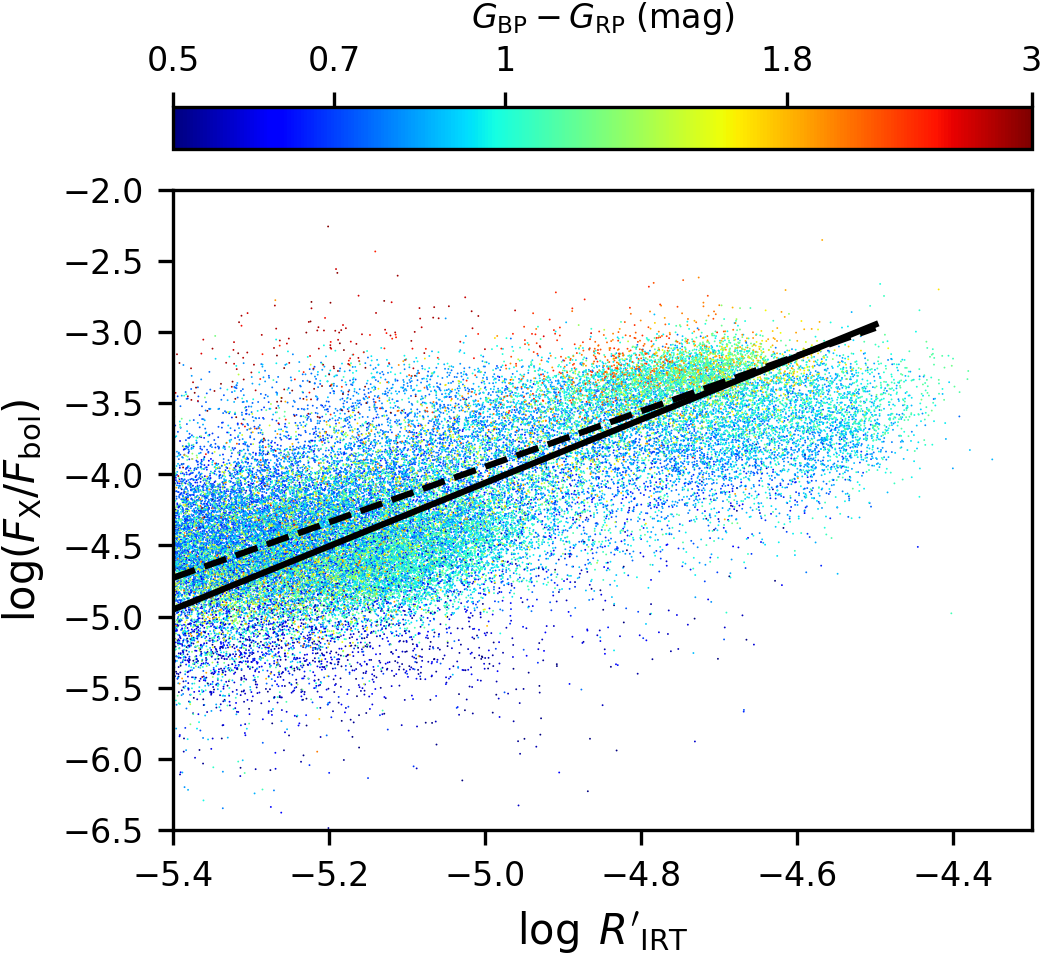}
		\caption{X-ray to bolometric flux ratio as function of the index $R^\prime_\mathrm{IRT}$ for the \textit{Gaia} DR3 sources with reliable activity index and likely eRASS:5 identification. The color scales with the logarithm of the $G_\mathrm{BP}-G_\mathrm{RP}$ color.
        The black dashed and solid lines represent the best fits to the eRASS1 and eRASS:5 data.}
		\label{fig: activity index F_X/F_bol vs. log_R}
	\end{figure}

    To investigate the relation between chromospheric and coronal
    activity, we show the X-ray to bolometric\footnote{We adopted the bolometric corrections from a table based upon \citet{pecaut13} and regularly updated under \url{http://www.pas.rochester.edu/~emamajek/EEM_dwarf_UBVIJHK_colors_Teff.txt} (current version 2022.04.16).} flux ratio as derived from eRASS:5 as a function of the $R^\prime_\mathrm{IRT}$-index as derived by \citet{lanzafame23}
	in Fig.~\ref{fig: activity index F_X/F_bol vs. log_R}.
    As evidenced by Fig.~\ref{fig: activity index F_X/F_bol vs. log_R},
    both activity indicators are reasonably well correlated for the highly and very highly active sources.
    We then fitted a linear relation between the logarithmic X-ray to bolometric flux ratio and the 
    logarithmic $R^\prime_\mathrm{IRT}$-index for the VHA and HA sources satisfying $\log(R^\prime_\mathrm{IRT})>-5.4$ using the ordinary least-squares (OLS) bisector regression as described by \citet{isobe90}. The resulting best fit parameters and the Pearson correlation coefficients (PCC) are listed in Table~\ref{tab: param bisector spectral type}, the best fit is also shown as black solid line in Fig.~\ref{fig: activity index F_X/F_bol vs. log_R}. For comparison, we show in Fig.~\ref{fig: activity index F_X/F_bol vs. log_R} the best fit to the eRASS1 detected sources as black dashed line. The slope of the fit is slightly flatter because more sources at low fractional X-ray fluxes remain undetected.

\begin{figure}[t]
		\includegraphics[width=\hsize]{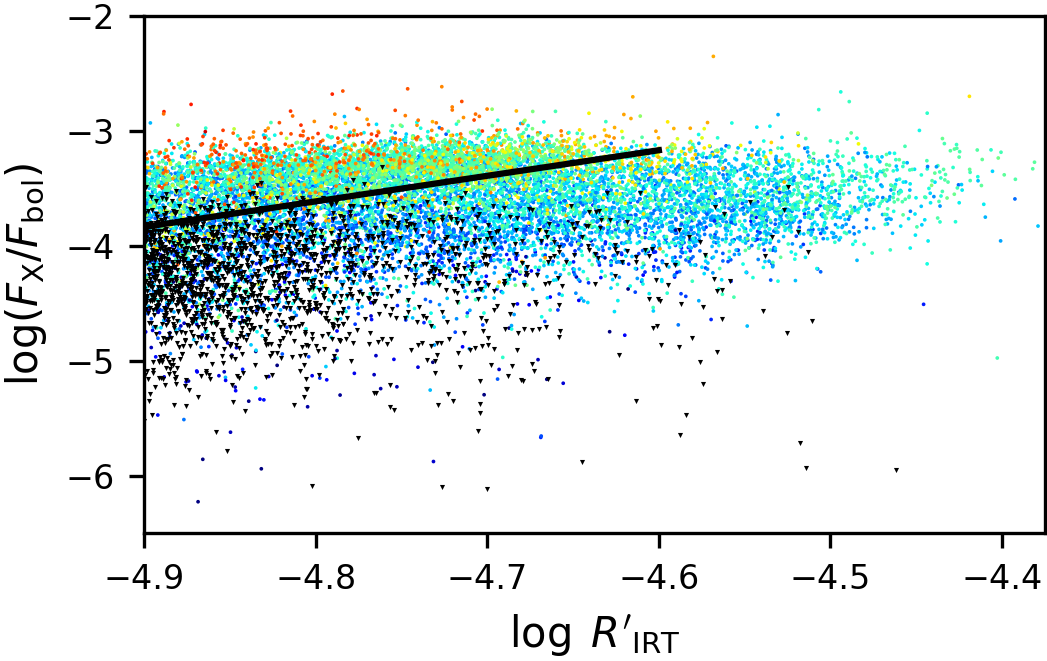}
		\caption{X-ray to bolometric flux ratio as function of the index $R^\prime_\mathrm{IRT}$ for the VHA sources. For the detected sources, the color scales with the logarithm of the $G_\mathrm{BP}-G_\mathrm{RP}$ color (same as in Fig.~\ref{fig: activity index F_X/F_bol vs. log_R}) and the upper limits are indicated by the black triangles. The black solid line represents the best fit of Fig.~\ref{fig: activity index F_X/F_bol vs. log_R}.}
		\label{fig: activity index F_X/F_bol vs. log_R vha upper limits}
	\end{figure}

    Although most of the VHA sources are also detected in X-rays, about 2\,500 of these remain undetected even in eRASS:5. The detection fraction of the VHA sources is the highest for K-type dwarfs, while almost half of F-type sources remain undetected in eRASS:5 (see Table~\ref{tab: eRASS:5 detection fraction});
    this group accounts for 53\% of the undetected VHA sources;
    we obtained upper limits for the undetected VHA sources from \citet{Tubin-Arenas24}. In some cases, the upper limits are less reliable due to a nearby eRASS:5 source. Although most of these X-ray sources are unlikely to be associated with the VHA source, they may prevent the detection of another nearby X-ray source. In Fig.~\ref{fig: activity index F_X/F_bol vs. log_R vha upper limits} we compare the upper limits of the remaining 1\,600 sources with the detected VHA sources. Even for sources with $\log(R^\prime_\mathrm{IRT})>-4.7$, we find some detected sources down to $\log(F_X/F_\mathrm{bol})=-4$ causing the correlation to appear quite flat but they are mostly F- and G-type sources that generally do not reach very high fractional X-ray fluxes (cf. Sect.~\ref{sec: Dependence on spectral type}). Most of the upper limits on the fractional X-ray flux are lower than for similar detected sources, which shows that the reason for their non-detection is a lower X-ray activity of these sources.

	\subsection{Dependence on spectral type}
    \label{sec: Dependence on spectral type}
\begin{figure}[t]
		\includegraphics[width=\hsize]{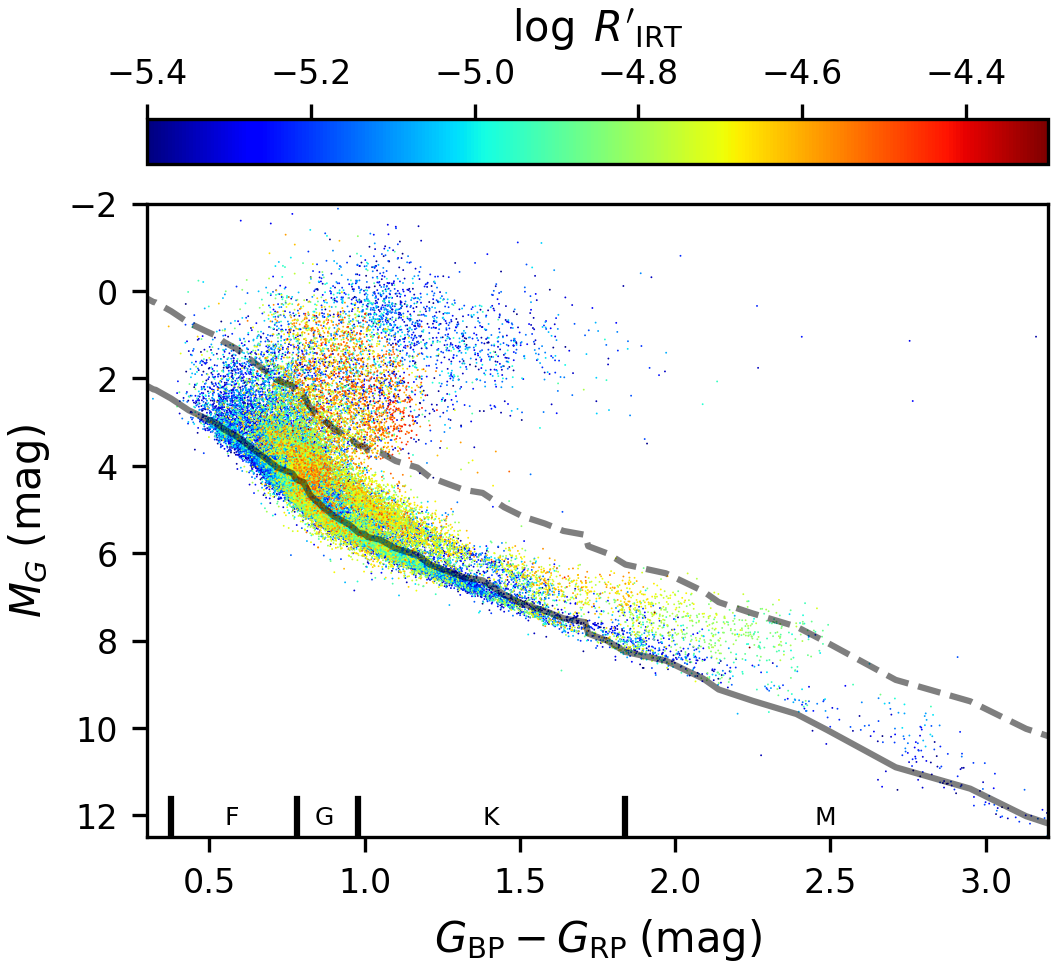}
		\caption{Color--magnitude diagram of the eRASS:5 sources with reliable activity index in \textit{Gaia} DR3. The color scales with the index $R^\prime_\mathrm{IRT}$. The solid line indicates the main sequence and sources located on the dashed line are 2~mag brighter than main sequence sources. }
		\label{fig: activity index HRD}
	\end{figure}

	The activity of the sample sources also varies for different spectral types as shown by a color--magnitude diagram of these sources in Fig.~\ref{fig: activity index HRD}. Most of the F-type sources have rather low chromospheric activity, while higher activity values are found among the later type stars. Especially for M-type dwarfs, sources with high and low activity are separated, with highly active dwarfs being located above the main sequence, which indicates youth.  Among giants, highly active sources are found in the subgiant regime, likely caused by RS CVn-type systems, while the more evolved stars are generally less active.

	\begin{figure*}[t]
        \sidecaption
		\includegraphics[width=12cm]{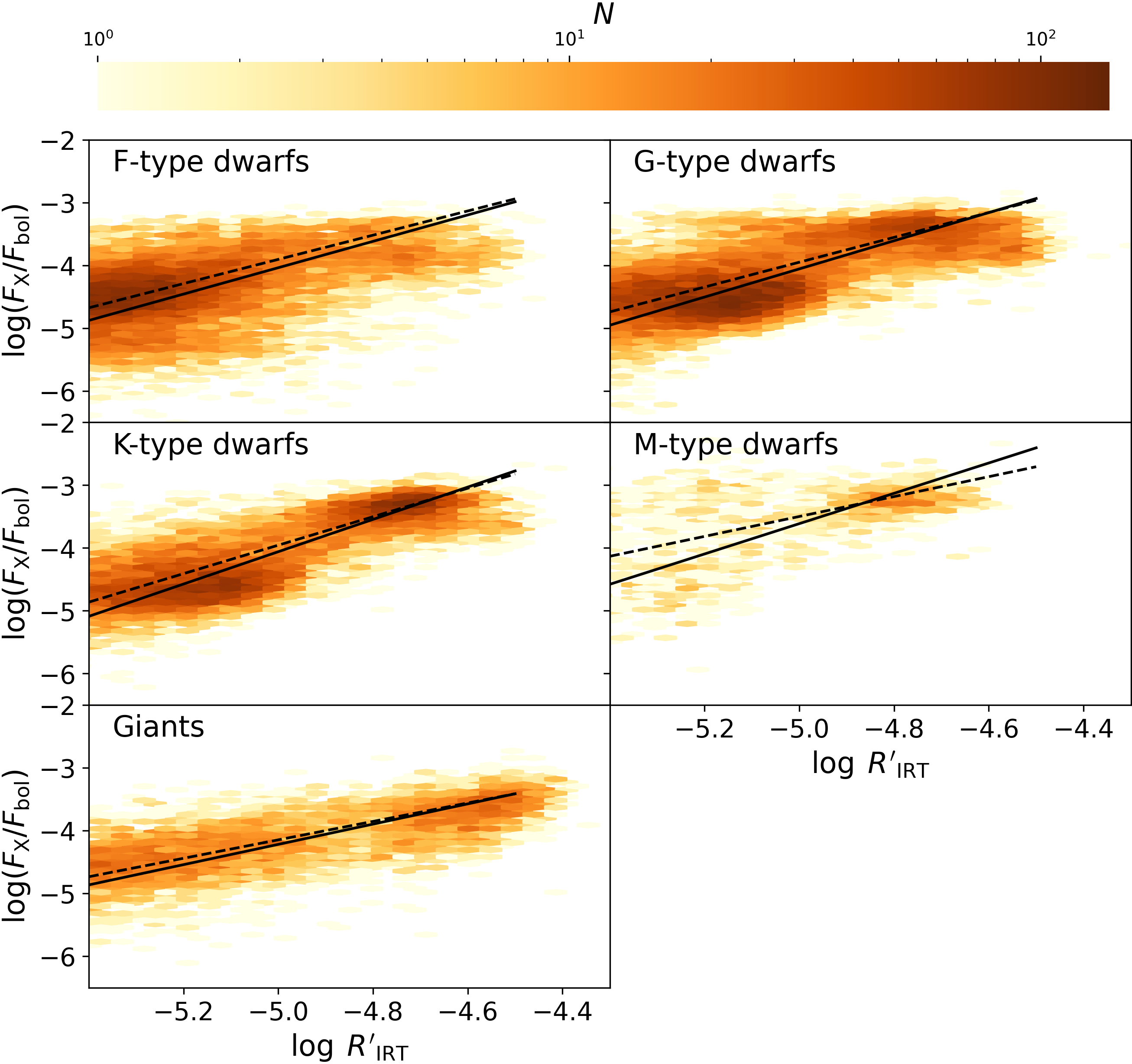}
		\caption{X-ray to bolometric flux ratio as function of the index $R^\prime_\mathrm{IRT}$ for likely eRASS:5 identifications with \textit{Gaia} DR3 sources of different spectral type (see text for details). The color scales with the number of sources within the bin and the black dashed and solid lines represent the best eRASS1 and eRASS:5 fits (cf. Table~\ref{tab: param bisector spectral type}).} 
		\label{fig: activity index spectral type}
	\end{figure*} 
    We compare the relation between the X-ray over bolometric flux ratio and the  $R^\prime_\mathrm{IRT}$-index for the different spectral types in Fig.~\ref{fig: activity index spectral type} and specify the best fit parameters and their statistical uncertainties of the OLS bisector regression for sources with $\log(R^\prime_\mathrm{IRT}) > -5.4$ in Table~\ref{tab: param bisector spectral type}. We note that, e.g., the extension of the fit to different $R^\prime_\mathrm{IRT}$-indices affects the parameters more strongly than the provided statistical errors. While both activity indicators are poorly correlated for F-type dwarfs, the strength of the correlation as well as the slope of the regression curve increase for G- and K-type stars. The slope and the correlation coefficient decrease again for M-type dwarfs because many of these sources in our sample are saturated in X-rays. We also find a strong correlation over a large range for giants although with a flatter slope. The best fits to the eRASS1 detected sources (shown by the black dashed lines in Fig.~\ref{fig: activity index spectral type}) are very similar except for M-type dwarfs, for which the number of sources is quite small especially for lower activities.
	\begin{table}[t]
		\caption{Parameters of OLS bisector for the different spectral types}
		\label{tab: param bisector spectral type}
		\centering
		\begin{tabular}{l r *{3}{l}}
			\hline \hline
			Sp type & number & slope & intercept & PCC \\
			\hline
			All & 43\,200 & 2.2240 $\pm$ 0.0074 & 7.060 $\pm$ 0.040 & 0.67 \\
			F & 11\,900 & 2.099 $\pm$ 0.017 & 6.455 $\pm$ 0.099 & 0.45 \\
			G & 14\,500 & 2.240 $\pm$ 0.014 & 7.144 $\pm$ 0.068 & 0.69 \\
			K & 10\,900 & 2.572 $\pm$ 0.017 & 8.801 $\pm$ 0.065 & 0.81\\
			M & 1\,100 & 2.414 $\pm$ 0.056 & 8.46 $\pm$ 0.36 & 0.57 \\
			Giants & 4\,800 & 1.612 $\pm$ 0.017 & 3.842 $\pm$ 0.082 & 0.71 \\
			\hline
		\end{tabular}
	\end{table}

	\subsection{Flux-flux relations} 
	\begin{figure*}[t]
        \sidecaption
        \includegraphics[width=12cm]{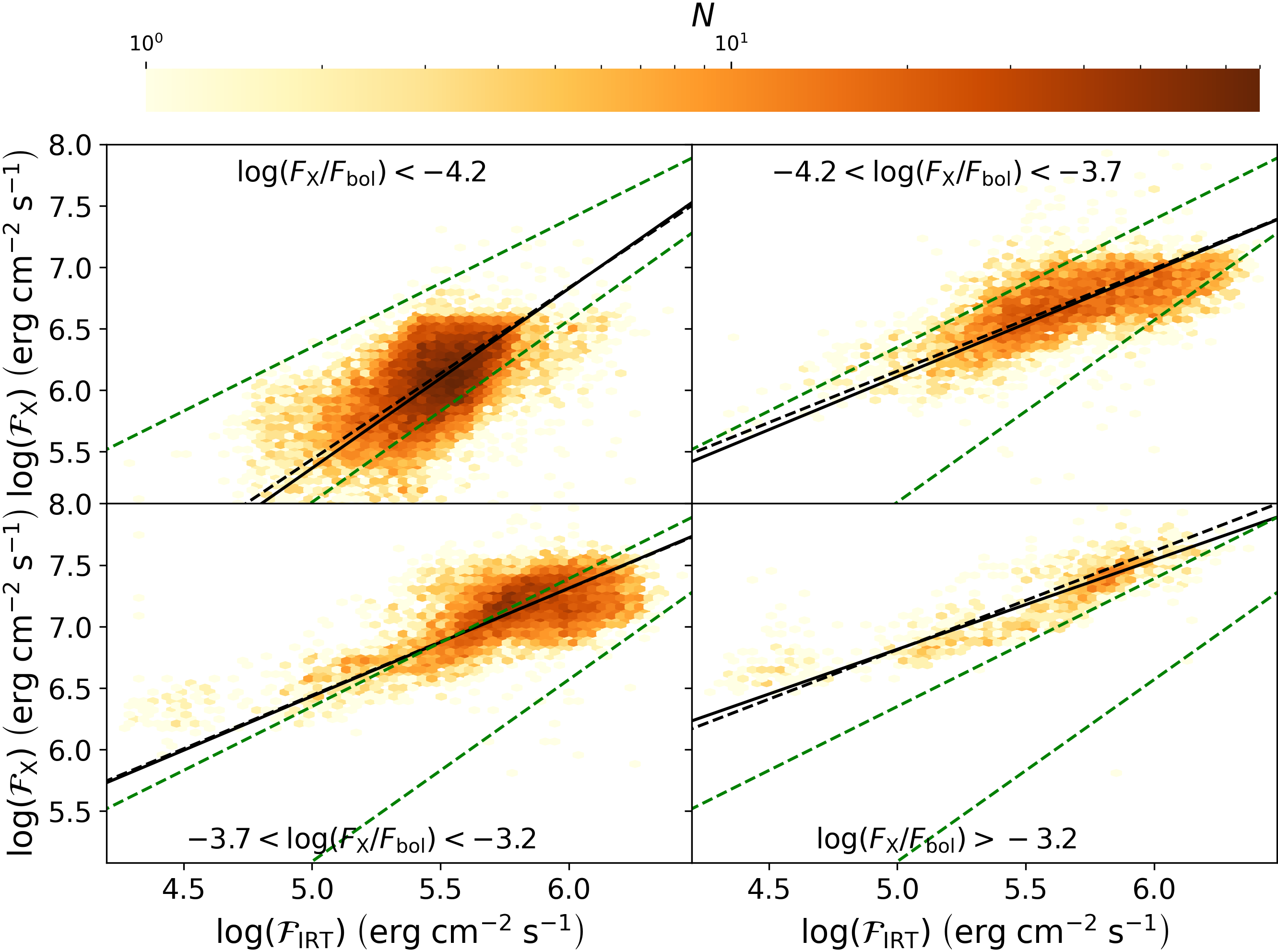}
		\caption{eRASS:5 X-ray as function of the IRT surface fluxes for different ranges of the fractional X-ray flux. The color scales with the number of sources within the bin, the best fits of our data are shown by the black slopes (dashed and solid lines for eRASS1 and eRASS:5 data) and the green dashed lines indicate the upper and lower branch for the data of \citet{martinez11} (cf. Table~\ref{tab: param bisector activity levels}).} 
		\label{fig: flux-flux relation}
	\end{figure*} 
	We also estimated the observed, mean surface flux of the IRT lines $\mathcal{F}_\mathrm{IRT}$ by Equation~\ref{equ: flux IRT} and the surface X-ray flux through 
	\begin{equation}
		\mathcal{F}_X = F_X \cdot \frac{d^2}{R^2}, 
	\end{equation}
	where $d$ is the distance estimated by the inverse parallax and $R$ is the radius of the  star provided by GSP-Phot. Since the radius is not available for a few sources, we have surface IRT and X-ray fluxes for 41\,300 eRASS:5 detetcted sources. 
	
	The flux-flux relation of the much smaller sample by \citet{martinez11} can be best described by two branches with the upper branch consisting of sources saturated in X-rays. Therefore, we divided our sample in different ranges of $\log(F_X/F_\mathrm{bol})$ and show the relations of the IRT and X-ray surface fluxes in Fig.~\ref{fig: flux-flux relation} and compare the best-fit parameters of the slopes in Table~\ref{tab: param bisector activity levels}. Since the correlation for F-type dwarfs is very weak in all samples, we excluded these sources from this analysis. 
    The flux-flux relation of our sources somewhat differs from the best fits of \citet{martinez11}. We also see a decreasing slope for sources with a higher fractional X-ray flux, but this is a more gradual development without two separate branches. Again, the differences between the best fits for the eRASS1 and eRASS:5 detected sources are small.
	\begin{table}[t]
		\caption{Parameters of OLS bisector for the different activity levels}
		\label{tab: param bisector activity levels}
		\centering
		\begin{tabular}{l r *{2}{l}}
			\hline \hline
			$\log(F_X/F_\mathrm{bol})$ & number & slope & intercept \\
			\hline
			$<-4.2$ & 15\,000 & 1.461 $\pm$ 0.012 & -1.939 $\pm$ 0.063 \\
			-4.2 -- -3.7 & 6\,000 & 0.863 $\pm$ 0.012 & 1.794 $\pm$ 0.064 \\
			-3.7 -- -3.2 & 7\,500 & 0.8791 $\pm$ 0.0097 & 2.038 $\pm$ 0.049 \\
			$>-3.2$ & 1\,100 & 0.727 $\pm$ 0.014 & 3.178 $\pm$ 0.078 \\
			lower branch\tablefootmark{a} & - & 1.48 $\pm$ 0.10 & -2.31 $\pm$ 0.54 \\
			upper branch\tablefootmark{a} & - & 1.04 $\pm$ 0.07 & 1.15 $\pm$ 0.43 \\
			\hline
		\end{tabular}
	\tablefoot{\tablefoottext{a}{\citet{martinez11}}}
	\end{table}

	\subsection{Correlation between the X-ray and Ca II H\&K fluxes}

\begin{figure}[t]
		\includegraphics[width=\hsize]{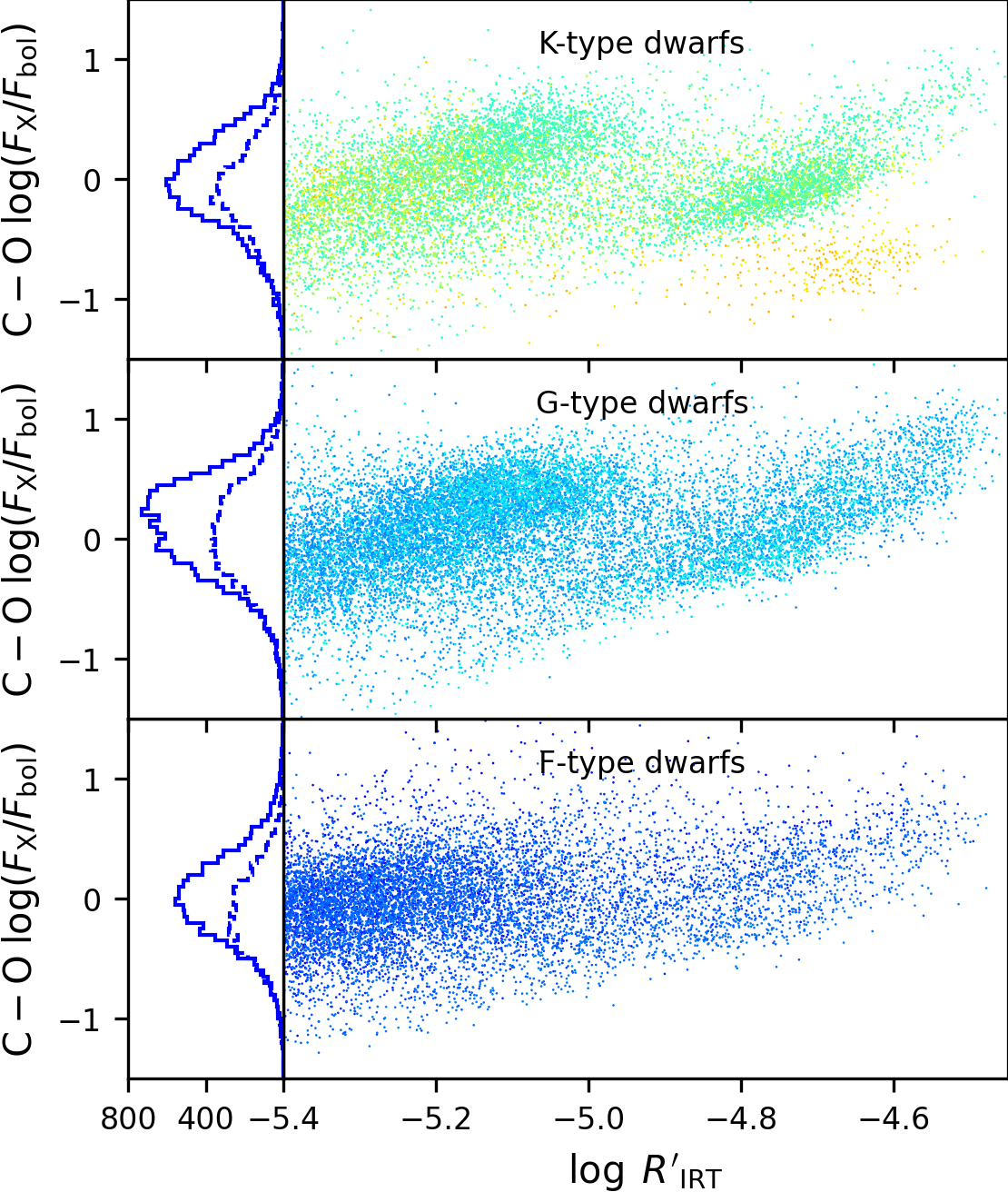}
		\caption{Calculated minus observed X-ray over bolometric flux values as a function of the $R^\prime_\mathrm{IRT}$-index. The panels show the different spectral types and the color coding of the sources is the same as in Fig.~\ref{fig: activity index F_X/F_bol vs. log_R}. The dots in the right panels show the values for the eRASS:5 detected sources and the dashed and solid histograms in the left panels show the distributions of the calculated minus observed values for eRASS1 and eRASS:5, respectively.}
		\label{fig: C-O F_X/F_bol}
	\end{figure}

    \citet{fuhrmeister22} studied the connection between the emission of the chromospheric Ca II H\&K lines and simultaneous X-ray measurements for 183 dwarfs and derived a correlation between the X-ray to bolometric flux ratio and the activity index $R^+_\mathrm{HK}$. To make their correlation comparable with our measurements of the X-ray and Ca~II IRT fluxes, we applied the conversion between the $R^\prime_\mathrm{IRT}$- and $R^\prime_\mathrm{HK}$-index as derived by Equation~9 of \citet{lanzafame23}. With these two correlations, we calculated the expected fractional X-ray fluxes from the $R^\prime_\mathrm{IRT}$-indices as provided in \textit{Gaia}~DR3. In Fig.~\ref{fig: C-O F_X/F_bol} we compare the calculated with the observed eRASS:5 fractional X-ray flux for the different spectral types. Since the sample of \citet{fuhrmeister22} contains only mid F- to late K-type dwarfs, their relation is only valid for these sources. In general, the X-ray activity is well predicted with a mean difference of 0.02~dex. However, the fractional X-ray flux of late K-type dwarfs is systematically underestimated. We also see a slight trend of increasing calculated fractional X-ray fluxes compared to the observed values with increasing $R^\prime_\mathrm{IRT}$-index, but also VHA sources at late spectral types are observed at higher X-ray activities. The standard deviation of the calculated minus observed values is with 0.40~dex larger than the error of 0.35~dex for the X-ray and Ca II H\&K fluxes reported by \citet{fuhrmeister22}, however, our observations are not simultaneous and can therefore be affected by  different states in the flaring activity and activity cycle.

    We note that \citet{fuhrmeister22} compared the fractional X-ray flux with the basal corrected H\&K index, while \citet{lanzafame23} applied the activity indices without basal correction, which might introduce a small bias. 
    Furthermore, the conversion also depends on the adopted relation between the IRT and H\&K activity indices. Applying other relations, the shape of the distribution remains similar, however, e.g., Equation~8 of \citet{lanzafame23} results in a general overestimation of the X-ray over bolometric flux by 0.50~dex, while the fractional X-ray flux is systematically underestimated by 0.47~dex if the relation from \citet{martin17} is adopted instead. With the eRASS1 sample, we observe similar but slightly higher fractional X-ray fluxes due to its lower sensitivity.

    \subsection{Relation with rotation}
    \label{sec: Relation with rotation}
    
	\begin{figure}[t]
		\includegraphics[width=\hsize]{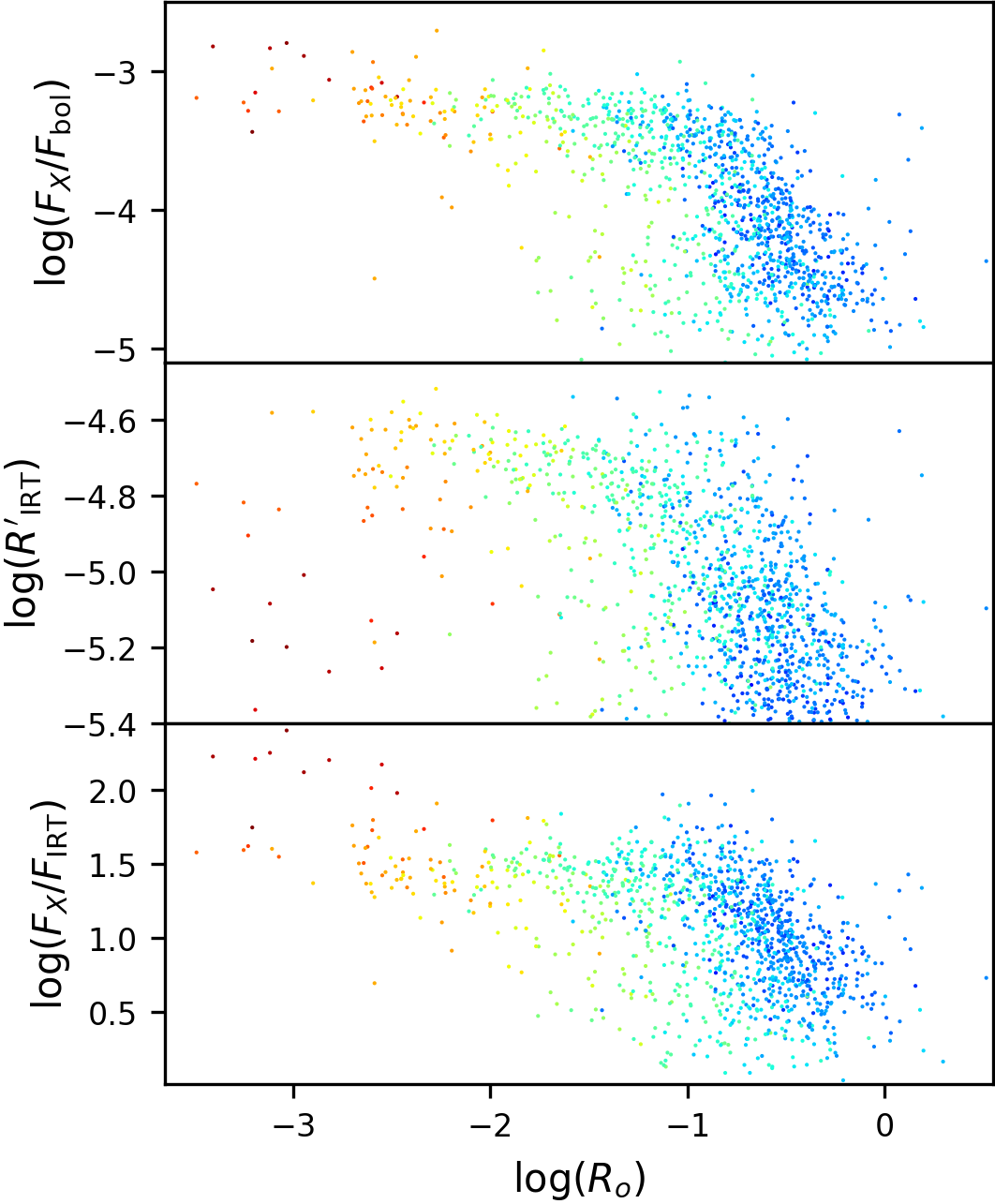}
		\caption{X-ray to bolometric flux ratio, $R^\prime_\mathrm{IRT}$-index, and X-ray over IRT flux ratio as a function of the Rossby number for the sources detected in eRASS:5. The color coding is the same as in Fig.~\ref{fig: activity index F_X/F_bol vs. log_R}.}
		\label{fig: Rossby F_X/F_IRT}
	\end{figure} 
    \citet{distefano23} provide rotation periods estimated from variability in \textit{Gaia} photometric time series. Rotation periods are available for 790 and 1\,200 of our sources with eRASS1 and eRASS:5 X-ray detection and a reliable $R^\prime_\mathrm{IRT}$-index and we show the different activity indicators vs. the Rossby number in Fig.~\ref{fig: Rossby F_X/F_IRT}, adopting the convection turnover times from the broken power law model from \citet{freund24}. For fast rotators, the X-ray and IRT emission saturate at $\log(F_\mathrm{X}/F_\mathrm{bol}) \approx -3$ and $\log(R^\prime_\mathrm{IRT}) \approx -4.5$ with the saturation regime starting at a similar Rossby number of $\log(R_O) \approx -1.4$. Hence, the ratio between X-ray and IRT flux is quite constant at a level of about $\log(F_\mathrm{X}/F_\mathrm{IRT}) = 1.5$. Late type stars often have a lower $R^\prime_\mathrm{IRT}$-index even in the saturation regime which might be caused by the transition from partially to fully convective stars or might indicate that the conversion between the activity index and $R^\prime_\mathrm{IRT}$-index (as estimated by Equation~7 of \citet{lanzafame23}) is inaccurate for these sources. For slow rotators, both activity indicators show a linear 
    decline, but the X-ray flux decreases faster than the IRT flux. This leads to a linear decrease of the flux ratio $\log(F_\mathrm{X}/F_\mathrm{IRT})$ although the scatter between the individual sources is quite large.

	\section{Conclusions}
    \label{sec: Conclusion}
    We combined Ca~II IRT measurements obtained from {\it Gaia} with X-ray fluxes obtained from eRASS1 and eRASS:5 and constructed a sample of 24\,300 and 43\,200 late-type stars, respectively, for which both activity indicators have been reliably detected. The X-ray detection fraction is largest for sources being very highly active in the IRT lines and for stars of spectral type K, where up to 90\% of the sources are detected in eRASS:5. The detection fraction very rapidly decreases for sources with lower IRT activity indices and F-type dwarfs often remain undetected as X-ray sources even if they are very active in the IRT lines.

    The $R^\prime_\mathrm{IRT}$-index generally shows a strong correlation with the X-ray to bolometric flux ratio, especially for K-type dwarfs and giants, while the two activity indicators are only poorly correlated for F-type dwarfs and sources in a low activity state. We emphasize that IRT activity values below $\log(R^\prime_\mathrm{IRT})<-5.4$ are not reliably detected in the current {\it Gaia} data release and should therefore be considered as inactive. As a consequence, no correlation with the X-ray activity is expected for these sources.
    For more active sources, the relation is well described by a linear slope but the slope increases towards later spectral types. When we compare the surface IRT and X-ray fluxes, we see a gradual change with the fractional X-ray flux but not two distinct branches as described by \citet{martinez11}. The deviation between the $R^\prime_\mathrm{IRT}$-index and the X-ray to bolometric flux ratio is with about 0.40~dex slightly larger than for simultaneously observed X-ray and Ca II H\&K fluxes \citep{fuhrmeister22}. This is probably caused by different activity regimes observed by \textit{Gaia} and \textit{eROSITA} observations due to the large time difference of the observations. When combined with rotation periods, we see that the X-ray to bolometric flux ratio and the $R^\prime_\mathrm{IRT}$-index saturates for fast rotating stars at a similar Rossby number causing the X-ray over IRT flux to be constant. For slow rotators, the fractional X-ray flux shows a steeper decline, and therefore, the X-ray to IRT flux ratio also decreases.

    \section*{Data availability}
    The catalog of the 24\,300 sources with reliable Ca II IRT activity and eRASS1 detection is available at the CDS via anonymous ftp to \url{cdsarc.cds.unistra.fr} (\url{130.79.128.5}) or via \url{https://cdsarc.cds.unistra.fr/viz-bin/cat/J/A+A/???}.
    Future \textit{eROSITA} data releases including eRASS:5 will be published following the timeline on \url{https://erosita.mpe.mpg.de/erass/}.

    \begin{acknowledgement}
    SF gratefully acknowledge supports through the Integrationsamt Hildesheim and the ZAV of Bundesagentur f\"ur Arbeit, SC by DFG under grant CZ 222/5-1, JR by DLR under grant 50 QR 2505, PCS by DLR under grants 50 OR 2215 and 50 OR 2412, and BF by DFG under program ID EI~409/20-1 and SCHN~1382/4-1. SF thanks Gabriele Uth and Maria Theresa Lehmann for their support.

    This work is based on data from eROSITA, the soft X-ray instrument aboard SRG, a joint Russian-German science mission supported by the Russian Space Agency (Roskosmos), in the interests of the Russian Academy of Sciences represented by its Space Research Institute (IKI), and the Deutsches Zentrum für Luft- und Raumfahrt (DLR). The SRG spacecraft was built by Lavochkin Association (NPOL) and its subcontractors, and is operated by NPOL with support from the Max Planck Institute for Extraterrestrial Physics (MPE).
    
    The development and construction of the eROSITA X-ray instrument was led by MPE, with contributions from the Dr. Karl Remeis Observatory Bamberg \& ECAP (FAU Erlangen-Nuernberg), the University of Hamburg Observatory, the Leibniz Institute for Astrophysics Potsdam (AIP), and the Institute for Astronomy and Astrophysics of the University of Tübingen, with the support of DLR and the Max Planck Society. The Argelander Institute for Astronomy of the University of Bonn and the Ludwig Maximilians Universität Munich also participated in the science preparation for eROSITA.
    
    The eROSITA data shown here were processed using the eSASS/NRTA software system developed by the German eROSITA consortium.

    This work has made use of data from the European Space Agency (ESA) mission
    {\it Gaia} (\url{https://www.cosmos.esa.int/gaia}), processed by the {\it Gaia}
    Data Processing and Analysis Consortium (DPAC,
    \url{https://www.cosmos.esa.int/web/gaia/dpac/consortium}). Funding for the DPAC
    has been provided by national institutions, in particular the institutions
    participating in the {\it Gaia} Multilateral Agreement.
    \end{acknowledgement}

	\bibliographystyle{aa} % style aa.bst
	\bibliography{mybib}

\end{document}